\documentclass[]{pasj01}

\begin{document} 
\Received{03-Jun-2020}
\Accepted{30-Oct-2020}

\title{Discovery of a new extreme changing-state quasar with 4 mag variation, SDSS J125809.31+351943.0}

\author{Shumpei \textsc{Nagoshi}\altaffilmark{1,}$^{*},$
Fumihide \textsc{Iwamuro}\altaffilmark{1},
Kazuma \textsc{Wada}\altaffilmark{1},
Tomoki \textsc{Saito}\altaffilmark{2}}

\altaffiltext{1}{Department of Astronomy, Kyoto University, Kyoto 606-8502, Japan}
\email{shumpei@kusastro.kyoto-u.ac.jp}

\altaffiltext{2}{Nishi-Harima Astronomical Observatory, Center for Astronomy, University of Hyogo, 407-2 Nishigaichi,
Sayo, Sayo, Hyogo 679-5313, Japan}


\maketitle

\begin{abstract}
We report the discovery of a quasar, SDSS J125809.31+351943.0 (J1258), which brightened in optical for 4 mag from 1983 to 2015, which is one of the largest quasar brightening events so far. The history of optical photometry data of this quasar from the Catalina Real-time Transient Survey and All Sky Automated Survey for Super Novae (ASAS-SN), mid-infrared photometry data from the WISE satellite, and the broad emission line (BEL) flux obtained by spectroscopy of Sloan Digital Sky Survey shows their significant increases between 2003 and 2015. Investigating its CFHT photometric observations in 1983 and the USNO-B catalog, which contains data in 1975 and 1969, we found that the source was 4 mag fainter before than the peak of the recent ASAS-SN photometry. From the history of these data, we identified J1258 as a new Changing-State Quasar (CSQ). We also performed follow-up spectroscopic observations in December 2018 and May 2019, using the 2-meter telescope in Nishi-Harima Astronomical Observatory. The results show that the continuum flux and the BEL flux decreased to about 50 \% of its peak. This indicates that J1258 is causing two changing-states for the BEL flux and continuum flux. We argue that J1258's variability, especially its brightening event, can be explained by the propagation of the heating front and the accretion disk state transitions based on the timescale and Eddington ratio variations. The estimated mass of the black hole of J1258 is about an order of magnitude larger than the CSQs found so far. Since both the changing timescale and the size of the accretion disk depend on the black hole mass, the J1258 brightening event can be interpreted as a scaled version of the variability in other CSQs. This suggests that samples of distant quasars with larger black hole masses may contain objects with longer and severer variations. 
\end{abstract}

\section{Introduction} \label{sec:intro}

The luminosity of quasars in various wavelength varies at various timescales. The typical scale of the variation in optical was thought a factor of $0.1$ mag on a timescale of a few days to months (e.g., \cite{2010ApJ...721.1014M}). The variation of the luminosity is usually interpreted as a result of local changes in the accretion disk, as phenomenologically explained in models, \textit{e.g}., the damped random walk (\cite{2010ApJ...721.1014M}). However, actual observations frequently show results that are inconsistent with popular theories or empirical rules. For example, the standard disk model predicts much longer viscous timescales than actually observed ones of variation. More puzzlingly, several recent large-scale optical surveys have shown that about a half of the quasars exhibit more than 1 mag of optical variability that cannot be explained by conventional models (\cite{2018yCat..18540160R}). The recent Sloan Digital Sky Survey (SDSS) provides a massive data set of multiple spectroscopic observations for millions of sources spanning $\sim$ 10 years. An interesting outcome from the survey is that some distant quasars show a so-called optical ``changing-look'' phenomenon (\cite{2015ApJ...800..144L}, \cite{2016MNRAS.457..389M}, \cite{2019ApJ...874....8M}, \cite{2016ApJ...826..188R}, \cite{2017ApJ...835..144G}, \cite{2018ApJ...858...49W}, \cite{2018ApJ...864...27S}, \cite{2019ApJ...883...94T}, \cite{2019arXiv190502904S}, \cite{2020ApJ...888...58G}) whereby broad emission lines (BELs) appear or disappear, which had been previously observed only in the Seyfert galaxies (\cite{1976ApJ...210L.117T}, \cite{1986ApJ...311..135C}, \cite{1989ApJ...337..236C}, \cite{1995ApJ...443..617S}). Most of the quasars that show a ``changing-look'' phenomenon, or Changing-Look Quasars (CLQ), show a common characteristic: large optical luminosity variations exceeding 1 mag. In recent years, such large luminosity changes have been found in many other objects besides CLQs, such as Hyper Variable Quasars (\cite{2018yCat..18540160R}) and Changing-State Quasars (CSQs; \cite{2020MNRAS.491.4925G}). The classification of the CLQ is based on the change of BELs, that of the HVQ is on the change of optical brightness, and that of the CSQ is on the change of BELs, optical continuum flux density, and mid-infrared luminosities. In this paper, we will refer to them collectively as CSQs. Objects that satisfy the CSQ classification criteria likely to satisfy the other two classification criteria as well.
 
Two classes of the origin are widely accepted as candidates to explain this type of variability: temporary shielding of the central region by optically thick material (\cite{2012ApJ...747L..33E}) and rapid changes in the mass accretion rate of the central engine (eg. \cite{2019ApJ...885...44D}, \cite{2016ApJ...826..188R}). Other factors such as micro-lensing (\cite{2017MNRAS.467.1259B}) and tidal disruption events (\cite{2015MNRAS.452...69M}) have also been proposed. In many cases in reality, change in mass accretion rate is believed to be the dominant cause on the basis of observation in X-ray (\cite{2016A&A...593L...9H}) and in optical polarization (\cite{2017A&A...604L...3H}). A main question is the origin and mechanism of such a significant change in the mass accretion rate and extreme change in the broad-line intensity that often accompanies the former.
 
Reverberation mapping is an effective way to investigate what is happening in the central region of the CSQ because it can estimate the structure (\cite{2014MNRAS.445.3073P}) of the broad line region (BLR) and determine the accurate black hole mass (\cite{2004ApJ...613..682P}). Reverberation mapping usually requires frequent spectroscopic obsrvations in every $\sim$10 days for a few years. Therefore, we explored a highly variable (more than 1 mag in optical) and relatively bright (less than 17 mag in optical) CSQ that would be feasible for reverberation mapping.

This paper describes the discovery of SDSS J125809.31+351943.0 (J1258) and discuss its large amplitude of variability. The method we used to find J1258 is described in Section \ref{sec:selection}. From Section \ref{sec:obs}, we focus on J1258. The data and observations that we used are explained in Section \ref{sec:obs}. Section \ref{sec:analysis} describes the analysis with which we find the object to show a variation in the optical and mid-infrared flux and identify it as a CSQ. Section \ref{sec:disussion} discusses the causes and timescales of the variation for 4 mag. The cosmological parameters are assumed to be $\rm H_0 = 70$ $
\rm kms^{-1}Mpc^{-1}$, $\rm \Omega_M = 0.3$ and $\rm \Omega_{\Delta} = 0.7$ based on the $\rm{\Lambda CDM}$ cosmology throughout this paper. 

\section{Sample selection} \label{sec:selection}
We performed a search for highly variable quasars, using the Catalina Real-time Transient Survey (CRTS) light curves in V band  and Sloan Digital Sky Survey (SDSS) quasar catalogue data release 7 (SDSS DR7Q; \cite{2010AJ....139.2360S}). The catalogue contains 105,783 quasars, and 19,547 of them are located in redshift of less than 0.8. We obtained CRTS light curves for 16,890 quasars among them to analyze their variation amplitudes and timescles.
\subsection{CRTS light curve}
The archival data for the CRTS (\cite{2009ApJ...696..870D}) consists of long-term $\sim$10 years) photometric observations obtained with three telescopes: the 0.7-m Cataline Sky Survey Schmidt (CSS) telescope, the 1.5-m Mount Lemmon Survey (MLS) telescope in Arizona, and the 0.5-m Siding Springs Survey (SSS) telescope in Australia. The survey observes an area of $2.5\times10^3$~deg$^2$ per night, with four 10-minute exposures per field of view. In addition, 21 observations were made for each lunar period on average. The observations were unfiltered and the measured flux values were converted into the Johnson V-band mag, with data for approximately 400 million objects at magnitude V $\rm{\sim}$ 20 and above and $\rm{Dec > -30}$ deg for the period 2003 to 2016 (CSS and MLS). In addition, the SSS provides data for the southern sky from 2005 to 2013 for about 100 million objects at magnitude $\rm{V\sim19}$ and above. 
\subsection{Light curve analysis}
\begin{figure}
\begin{center}
\includegraphics[width=80mm]{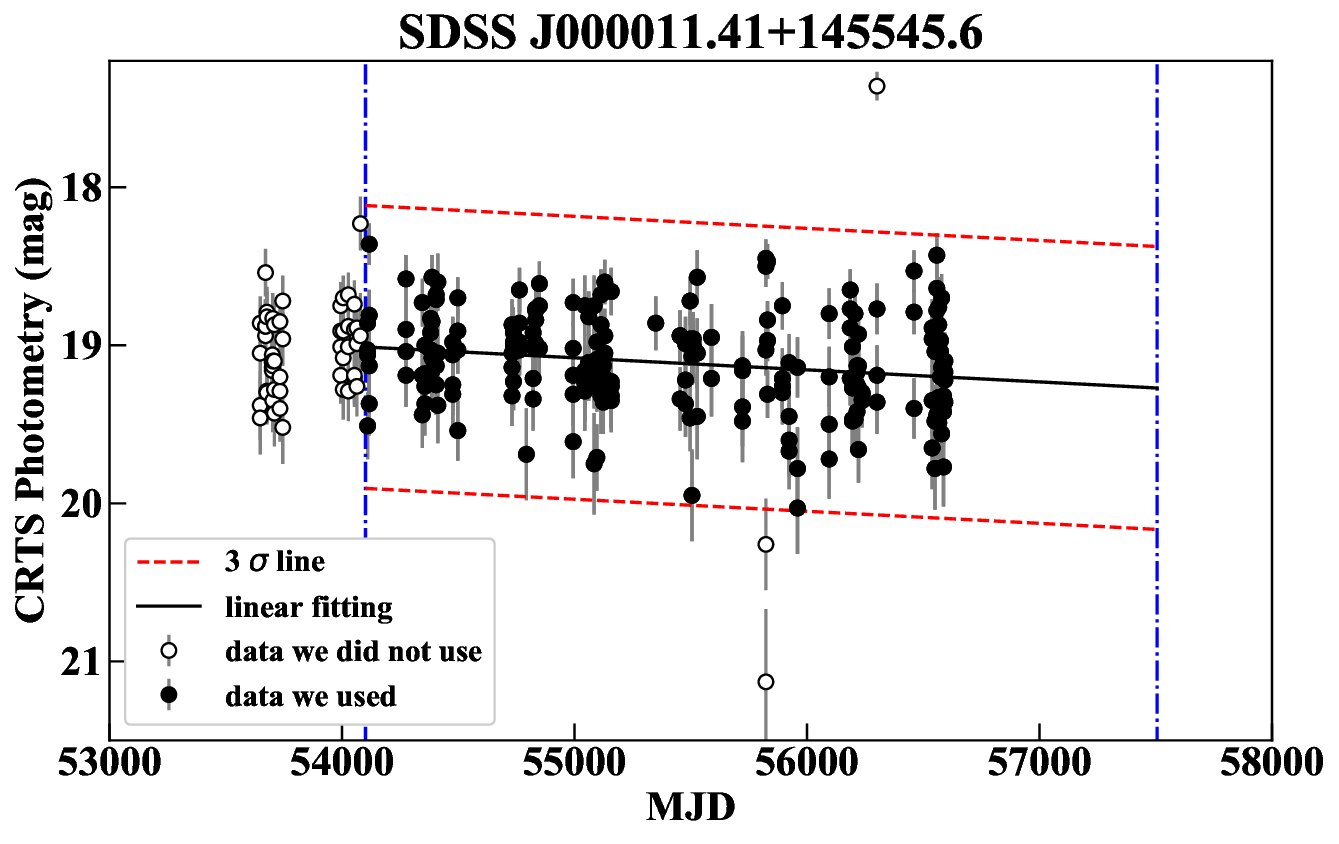}
 \end{center}
 \caption{An example of fitting a CRTS light curve. The black and white points represent the original catalog data. The blue dashed line represents the MJD 54101 and MJD 57506, and only the data between these line are} used. The solid black line is a fitting straight line and the dashed red line represents 3 $\sigma$.
 \label{fig:fitting_example}
\end{figure}

\begin{figure}
 \begin{center}
 \includegraphics[width=80mm]{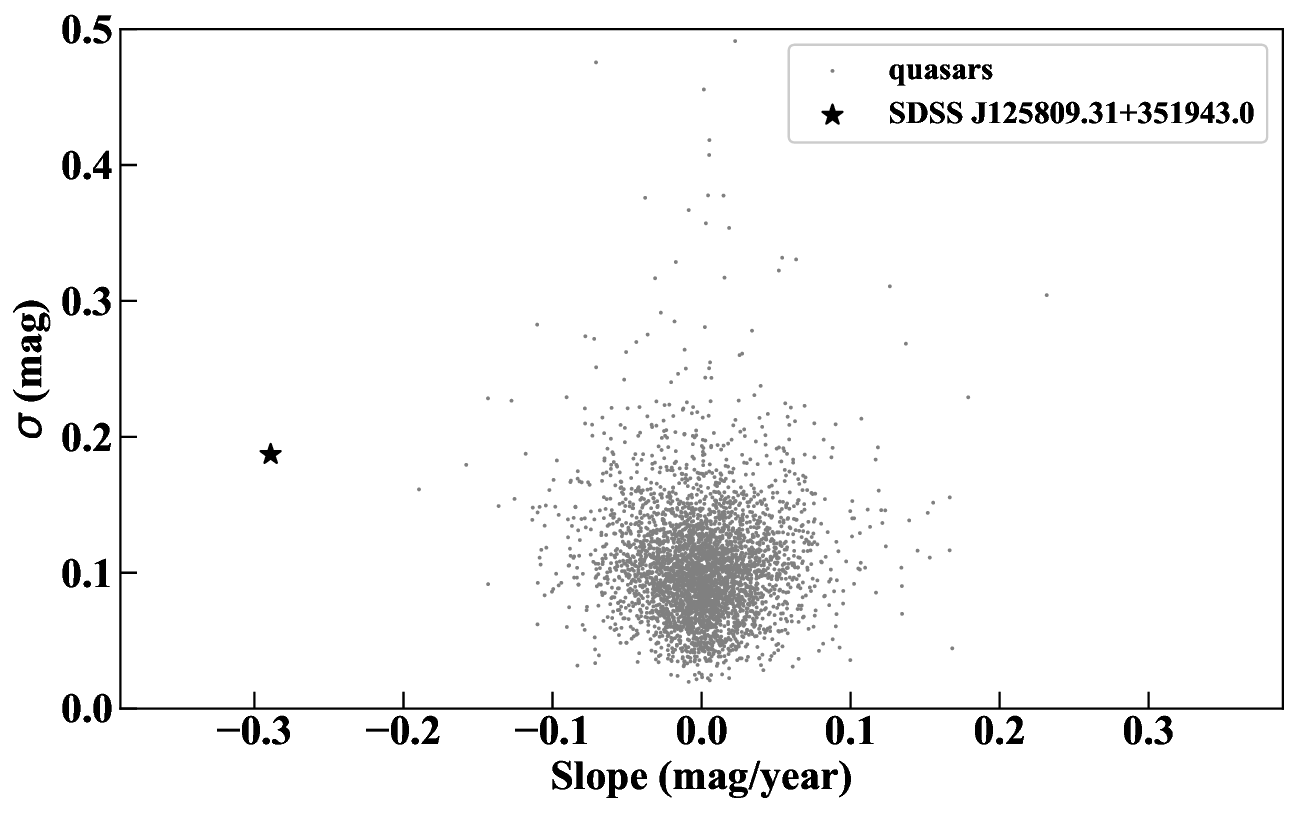}
 \end{center}
 \caption{A diagram summarizing the slope and $\sigma$ obtained from the fitting. The gray points represent the results of each fitting, and the star-shaped point represents SDSS 125809.31+351943.0.
 }
 \label{fig:slope_sigma}
\end{figure}

\begin{figure}
 \begin{center}
 \includegraphics[width=80mm]{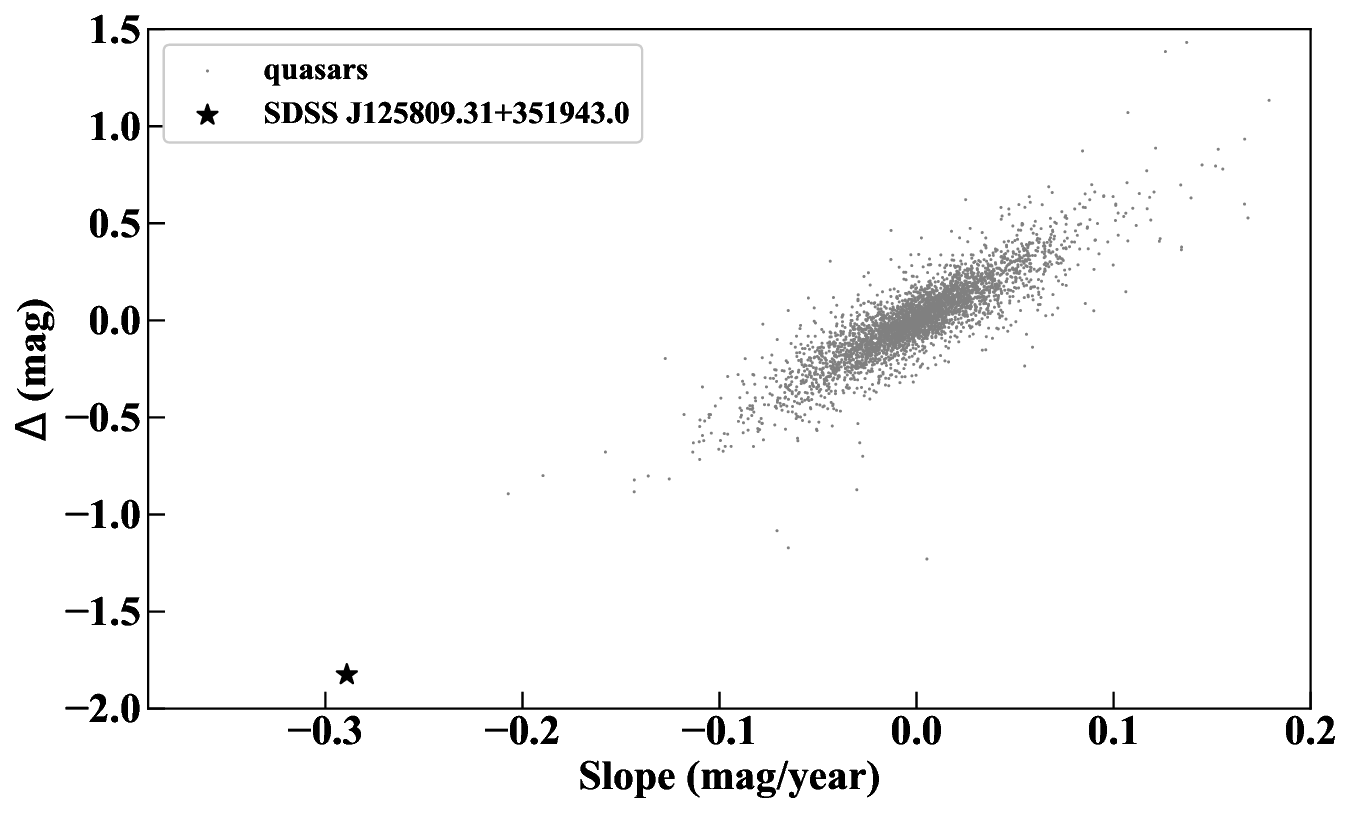}
 \end{center}
 \caption{A diagram summarizing the slope and $\Delta$ obtained from the fitting. The y-axis represents the amount of increment in the CRTS light curve from the first magnitude (the average of values measured within 100 days of the beginning of the light curve) to the last magnitude (the average of values measured within 100 days of the end of the light curve)). The gray points represent the result of each fitting, and the star-shaped point corresponds to SDSS 125809.31+351943.0.
 }
 \label{fig:slope_delta}
\end{figure}

Most quasars have random variations on short timescales, but CSQs tend to have monotonic variations on long timescales of a few years (e.g, \cite{2020MNRAS.491.4925G}). In order to find such monotonically variable objects, a linear function was used to fit the light curves using the least squares method. We used only the values in the MJD range from 54101 to 57506, because photometry before MJD 54101 can lead to incorrect values due to inappropriate aperture size (\cite{2019MNRAS.482...98D}). Values outside the $3\sigma$ ($\sigma$ is the standard deviation of the fitting) were excluded as outliers and re-fitted ($3\sigma$ clipping). This process was repeated until no outliers were found. An example of the fitting is shown in figure \ref{fig:fitting_example}. Such fittings were performed for all the light curves and narrowed down to 3,792 objects that included a duration of more than 1000 days and had an average CRTS's V-band magnitude brighter than 18. To find bright and variable objects for the following spectroscopic observation, we focused on those with slope of less than -0.1 mag/year of the fitted line and small $\sigma$ of the fitting. The distribution of the slope and $\sigma$ obtained by the fitting is shown in figure \ref{fig:slope_sigma}. The distribution of the slope and the optical magnitude differences between the first and last measurements ($\Delta$) is shown in figure \ref{fig:slope_delta}. In these diagrams, it can be seen that the variation of J1258 (the redshift is 0.31) is extremely larger than that of the other quasars.
 
 This method is similar to that used by \cite{2020MNRAS.491.4925G} (the primary difference is that they used Bayesian blocks to analyze light curves), but J1258 was not on their list. A comparison of our selection approach with that used in \cite{2020MNRAS.491.4925G} is discussed in section \ref{comparison}. From next section, we focus on J1258.

\section{Data and Observation} \label{sec:obs}

\subsection{Photometric data}
\subsubsection{ASAS-SN}
\begin{figure}
 \begin{center}
 \includegraphics[width=80mm]{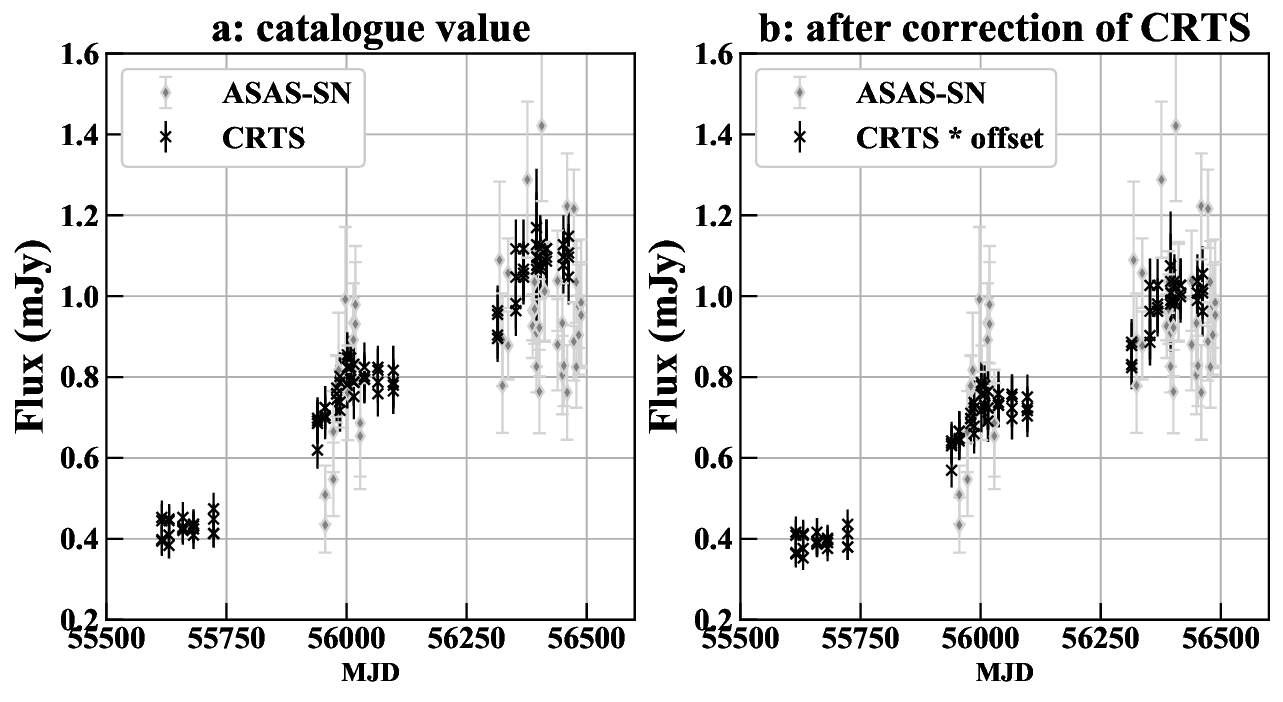}
 \end{center}
 \caption{
 (Left panel: a) cataloged (crosses) CRTS and (diamonds) ASAS-SN fluxes (mJy) for regions encompassing J1258 where both the observations were available. (Right panel: b) Same as panel b but with the CRTS fluxes replaced with the corrected ones. 
 }
 \label{fig:crts_calib}
\end{figure}

To investigate J1258's recent variability, we used the photometric data from the All Sky Automated Survey for Super Novae (ASAS-SN; \cite{2017PASP..129j4502K}), a project to automatically perform an all-sky survey, using 24 telescopes every observable day. The limiting magnitude of the ASAS-SN is about 17 mag. In the ASAS-SN, two filters, g-band and V-band, were used. We used the data of either filter for our analysis partly because the CRTS observations were made with no filters and majorly because a small difference between the data with the two filters is negligible compared with the variation of the target objects. There is a systematic discrepancy between the CRTS and ASAS-SN photometric systems. We corrected the filterless CRTS flux, using the ASAS-SN photometric data from the period when both observations were made, in the following procedure. First, for each CRTS photometric value, we calculated the average ASAS-SN photometric value within 20 days of the epoch when the CRTS data was taken. Next, we calculated the ratio of the CRTS photometric value to the average ASAS-SN value. Finally, we determined the offset by averaging the calculated ratios; the ASAS-SN flux is 0.87 times as bright as the CRTS flux (figure \ref{fig:crts_calib}).
 
 Then, outliers were removed from the ASAS-SN data, using linear fitting with a $3\sigma$ clipping iteration, applied to the year-by-year data separately. 
Figure \ref{fig:curve_old_fix} shows the light-curves of the cataloged data before and after the correction. 

\subsubsection{WISE}
The Wide-field Infrared Survey Explorer (WISE; \cite{2010AJ....140.1868W}), which is a mid-infrared satellite launched in December 2009 and observed the entire sky in four bands (3.4, 4.6, 12, and 22 $\rm{\mu m}$), also observed J1258. WISE operations were temporarily terminated in February 2011, but observations resumed in December 2013 as the NEOWISE project(\cite{2014ApJ...792...30M}). NEOWISE has two bands of 3.4 and 4.6 $\rm{\mu m}$ and observed the entire sky, visiting each area in the sky about ten times every six months. 

As the information of J1258, we referred to the single-exposure profile-fit magnitude within 3'' from RA = 12:58:09.31 and DEC = +35:19:43.03. From these values, we selected those with no contamination and confusion flag (``cc$\_$flags'' = 0000). In the present analysis, we used the combined data with each bin accumulated for six months period, sufficient for the variation timescale of our interest. 

\subsubsection{Before CRTS: USNO and CFHT}
\begin{figure}
\begin{center}
\includegraphics[width=80mm, clip]{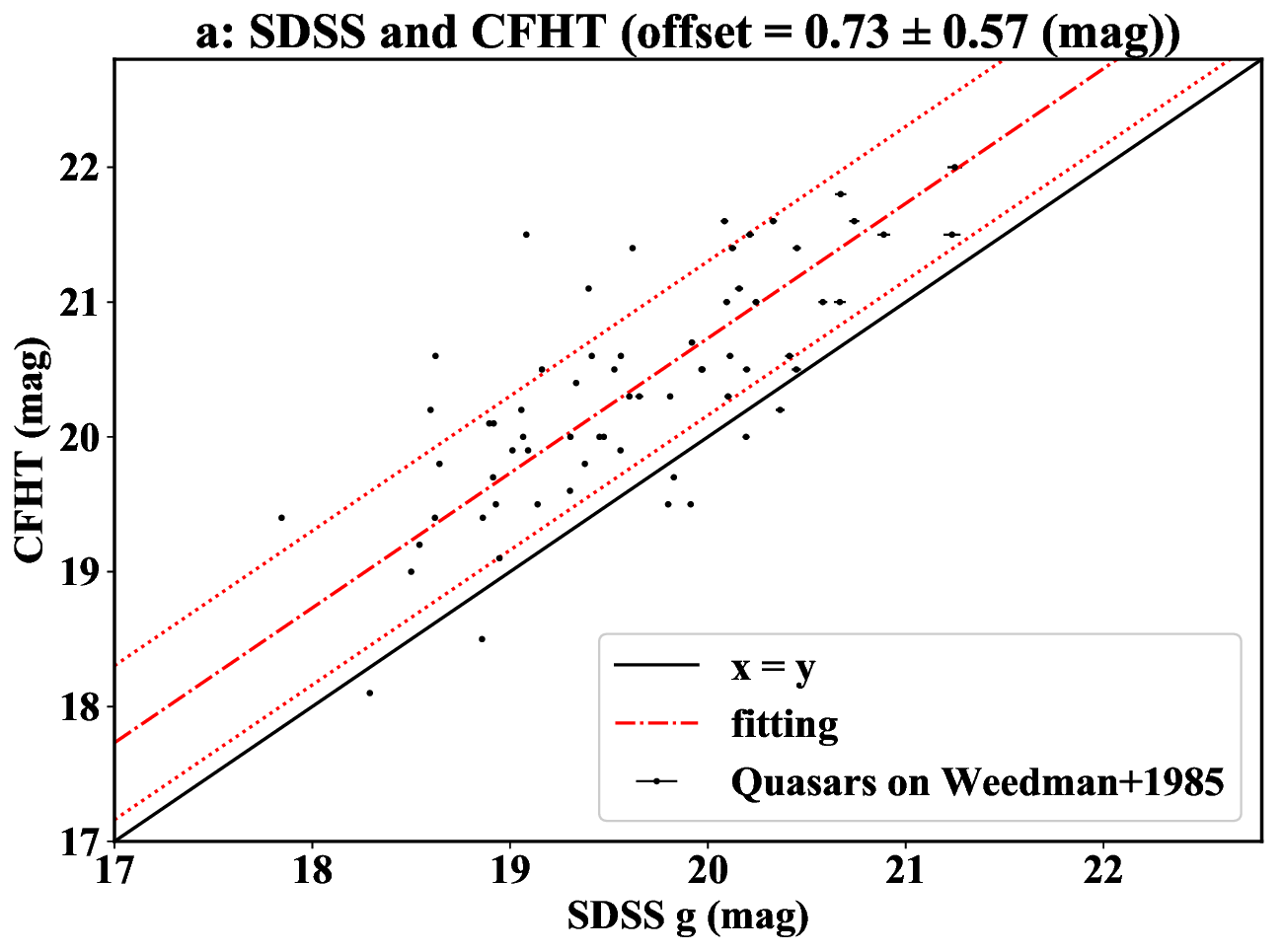}
\includegraphics[width=80mm, clip]{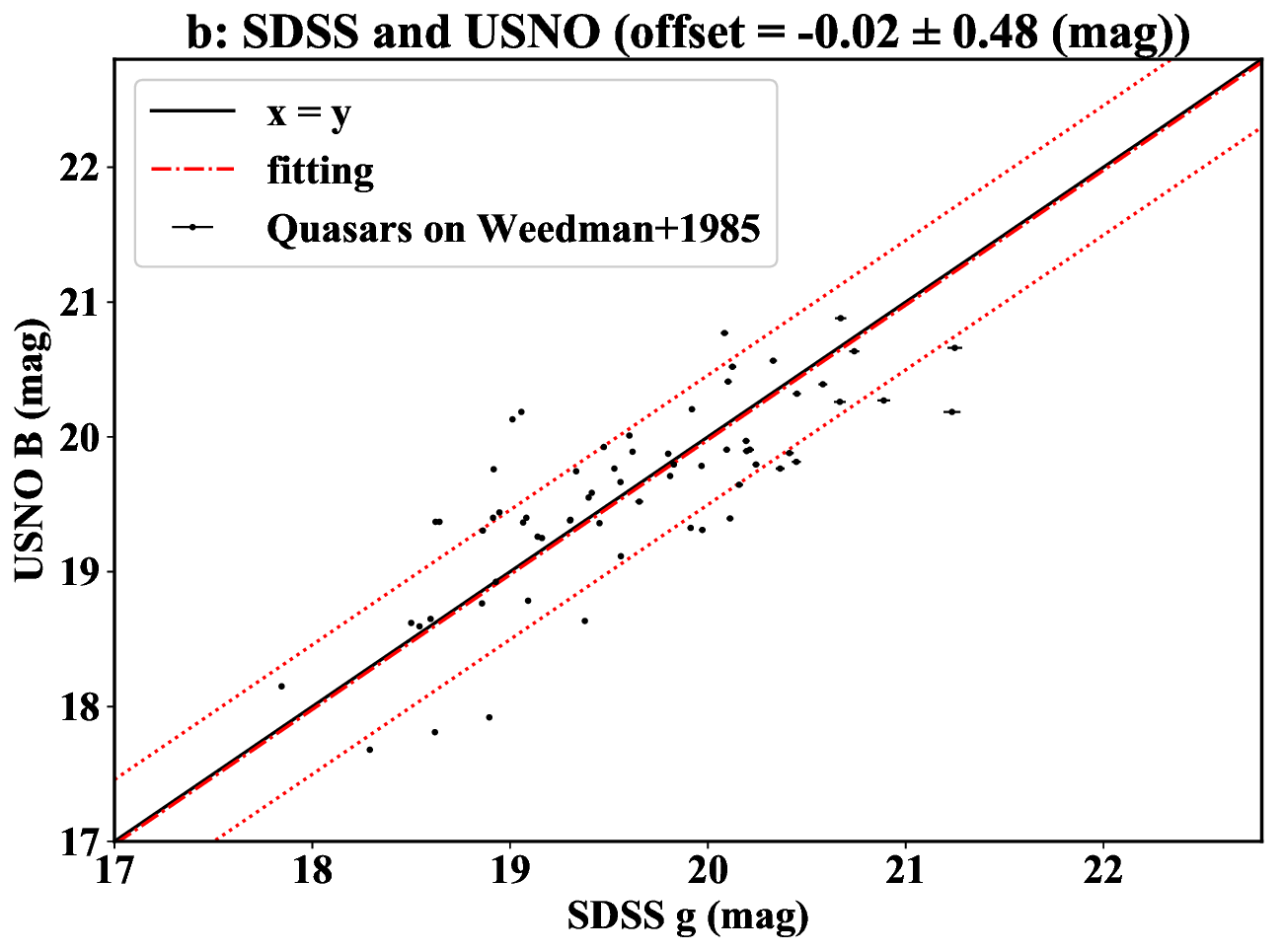}
\end{center}
\caption{
 (Top panel: a) Scatter plot of the photometric values of the SDSS g-band and CFHT of quasars listed in \cite{1985ApJS...57..523W}, and (bottom: b) those of the SDSS g-band and USNO-B catalog. The data points are taken from \cite{1985ApJS...57..523W}, the SDSS quasar catalog, and the USNO-B catalog. Black solid lines show $y=x$, and red dashed and dotted lines show the best-fit result with a linear function ($y=x+$const) and 1-$\sigma$ errors.
\label{fig:USNO_SDSS}}
\end{figure}

We also investigated photometric observations of  SDSS J125809.31+351943 before the era of the CRTS. We obtained archival data of photometric observations by the Canada-France-Hawaii Telescope (CFHT) in 1983 and by the United States Naval Observatory (USNO) in 1975 and 1969. No errors for the data are listed in the catalogs; however, given that their photometric observations were carried out with photographic plates, the errors are expected to be considerable. We estimated the systematic and measurement errors of the CFHT and USNO data, comparing their data of 189 AGNs reported in \cite{1985ApJS...57..523W}, which is a reference for the photometry in CFHT, with those of the $g$-band data in the SDSS DR12 Photometry Catalog (\cite{2015ApJS..219...12A}). Here, the comparison was made only with the quasars included in the SDSS DR12 quasar catalog to avoid mismatches with nearby stars. Figure \ref{fig:USNO_SDSS} compares the photometric values of the SDSS $g$-band with those of the CFHT, using 65 matched sources. We found a systematic offset of minus 0.686 mag with a typical error of 0.776 mag in the CFHT photometric observations. The same analysis was performed with regard to the USNO-B catalog values (figure \ref{fig:USNO_SDSS}), where the B-band of USNO is equivalent to the g-band of the SDSS. 

\begin{figure*}
 \begin{center}
 \includegraphics[width=160mm]{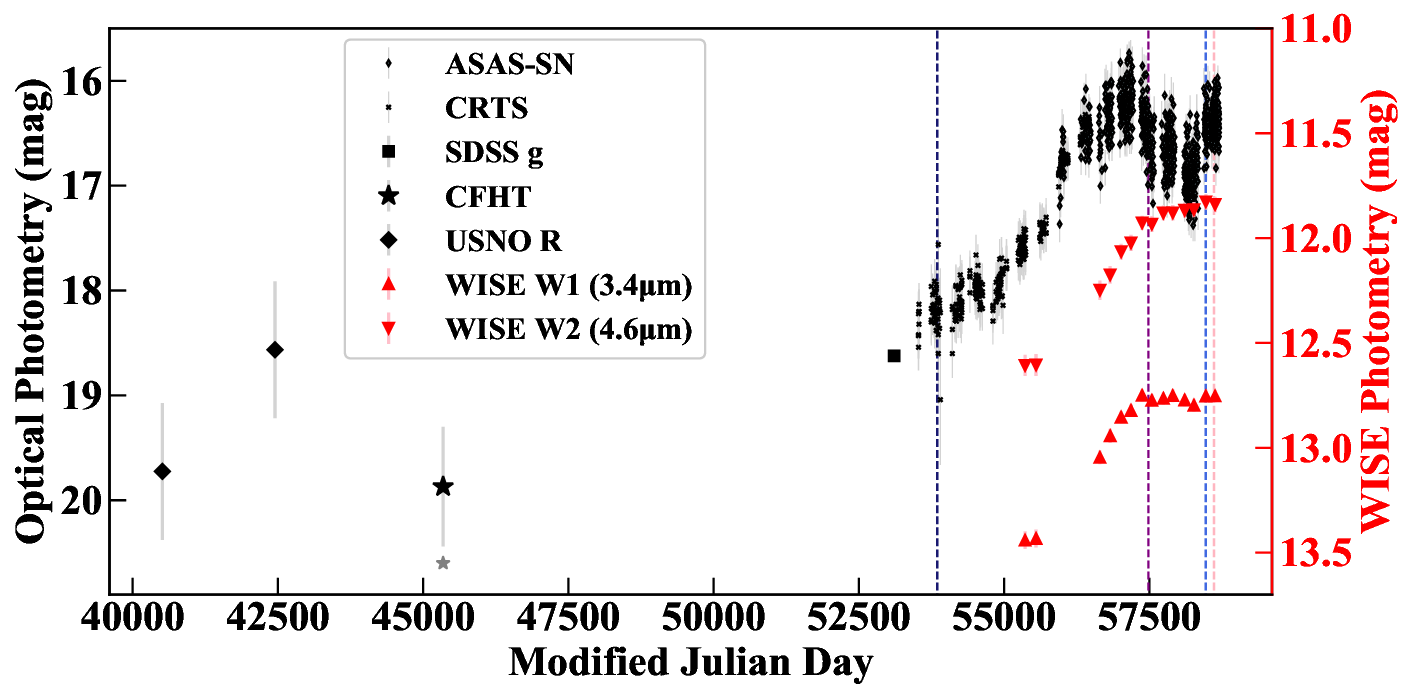}

 \end{center}
 \caption{
 Time-series of the photometric data of J1258 in optical and mid-infrared. Red triangles denote the photometric values of WISE in the scale indicated on the right vertical axis. The other marks (black) denote the optical broad-band flux density in the scale indicated on the left vertical axis. The plotted USNO and CFHT values are already corrected as in figure \ref{fig:USNO_SDSS}. The gray star-shaped point at around MJD=45000 indicates an uncorrected value for reference. The four vertical dashed lines indicate the epochs when the spectra of the source were taken with spectroscopic observations in the colors corresponding to those in figure \ref{fig:spectra}.
}
\label{fig:curve_old_fix}
\end{figure*}

Figure \ref{fig:curve_old_fix} summarizes all the photometric data of J1258 accompanied with the estimated errors, where the offsets are corrected. The result indicates that the luminosity of the object varied by 4 mag during 1983 to 2016. We should note that this object is radio quiet and hence that the jet contribution to this variability should be insignificant.

\subsection{Spectroscopic data and observation}
J1258 has been spectroscopically observed by the SDSS twice so far, in April 2006 (referred to as the first) and April 2016 (second). The SDSS (\cite{2000AJ....120.1579Y}) is based on a photometric survey using a 2.5-m wide-field telescope (\cite{2006AJ....131.2332G}) and 30 $\rm{2k \times 2k}$ CCDs and is followed by spectroscopic observations. Between the two observations, the spectroscopic instrument was upgraded, the wavelength range was extended, and the thickness of the fiber on the SDSS plate was changed from 3$^{\prime\prime}$ to 2$^{\prime\prime}$. 

We performed two new spectroscopic follow-up observations in December 2018 and May 2019 with the slit spectrograph called ``MALLS'' installed on the 2-meter telescope in Nishi Harima Astronomical Observatory in Japan. The grating was 150 /mm, giving a spectral resolution of $\sim 600$, with the $1.\!\!^{\prime\prime}2$ slit width and GG475 order cut filter. We executed five 1200-second exposures for two nights. The observed data were reduced with the standard processing of slit spectroscopy with IRAF (dark subtraction, flat correction, matching sky subtraction, wavelength correction, and flux correction using the spectrophotometric standard). 

\begin{figure}
 \begin{center}
\includegraphics[width=80mm]{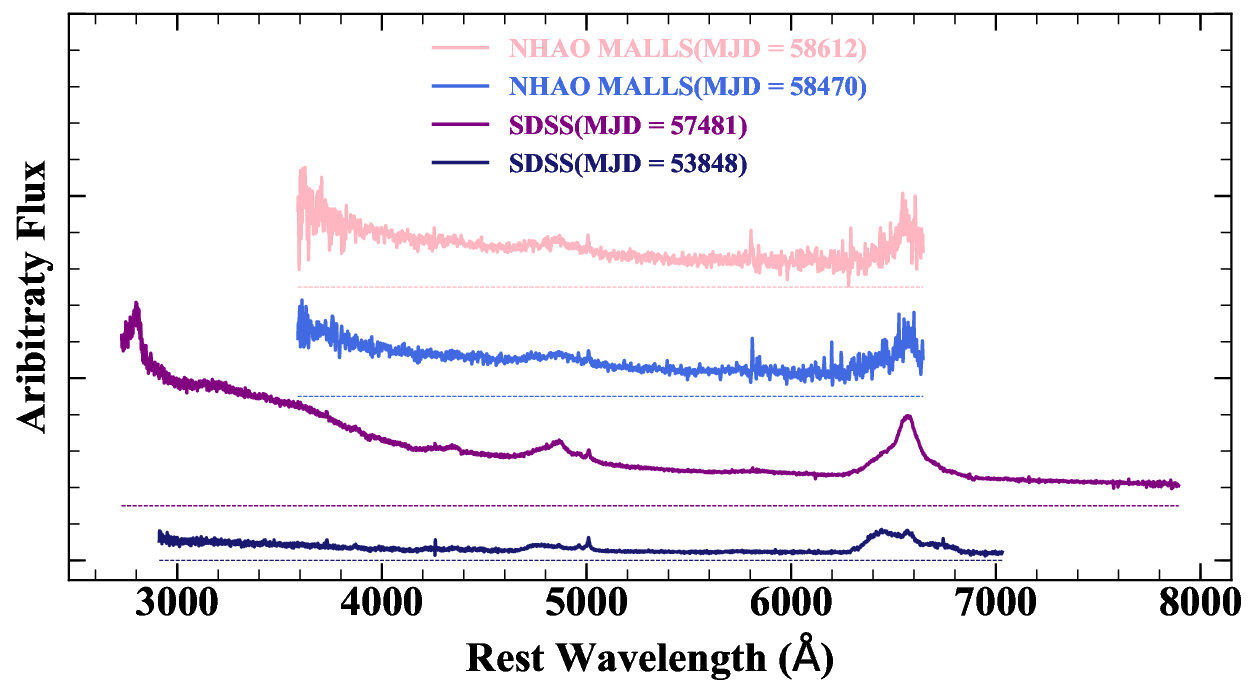}
\end{center}
 \caption{
 Spectra of J1258. From the bottom, the first SDSS spectrum (in dark-blue, MJD=53848), the second SDSS spectrum (purple, MJD=57481), the first MALLS spectrum (blue, MJD=58470), and the second MALLS spectrum (pink, MJD=58612). Each spectrum is normalized by the flux of [OIII]5007, and a constant is added to the flux density ($y$-axis) to make it easier to see. The zero point of each spectrum is drawn with a dotted line in the corresponding color.
 }
 \label{fig:spectra}
\end{figure}

\section{Analysis and Identification} \label{sec:analysis}
 The definition of the CSQ by \cite{2020MNRAS.491.4925G} with regard to its physical properties is summarized as follows. 1. it is not a blazer; 2. its optical luminosity varies greatly on a longer timescale than a typical quasar (A change of more than 1 mag satisfies the conditions, although the exact criteria is different. See the original paper for detailed conditions.); 3. it has more than 0.2 mag of variation in mid-infrared flux; 4. it has been spectroscopically observed twice or more, and the time-variation of the H$\rm{\beta}$/[OIII] ratio in its spectrum is more than 30\%.

 First, J1258 is not a known blazer. Also, according to the VLA/FIRST survey catalogue\footnote{http://sundog.stsci.edu/cgi-bin/searchfirst} (\cite{1995ApJ...450..559B}), no radio sources at 1365 or 1435 MHz were found within 30'' around the coordinates of J1258 (The detection limit of the survey is about 1 mJy). Here, We estimate the upper limit of the radio (about 1.4 GHz) to optical (about 550 THz) flux ratio ($R_{\rm r-o} \equiv f_\nu ({\rm radio}) / f_\nu ({\rm optical})$) as an indicator of its radio loudness (\cite{1989AJ.....98.1195K}). J1258 at its faintest time during the VLA/FIRST observation period (1993 to 2014) was about 19 mag (about 0.1 mJy) in V-band. Then, the upper limit of $R_{\rm r-o}$ is estimated as $R_{\rm r-o} < 10$. \cite{1989AJ.....98.1195K} classified samples which have R larger than 10 as radio-loud quasar. Therefor, J1258 is unlikely to be a radio-loud quasar.
 
Besides, the amount of the change of its optical luminosity is at least 2 mag in the last 15 years and 4 mag in a longer period. Furthermore, the W1 and W2 fluxes changed by more than 1 mag in WISE. Therefore, J1258 satisfies at least three out of the four criteria to be identified as a CSQ. The final piece of information to consolidate the identification is the ratio of H$\rm{\beta}$/[OIII]. We measured the change in the flux of H$\rm{\beta}$ in the following procedure.

\begin{figure}
\begin{center}
\includegraphics[width=80mm, clip]{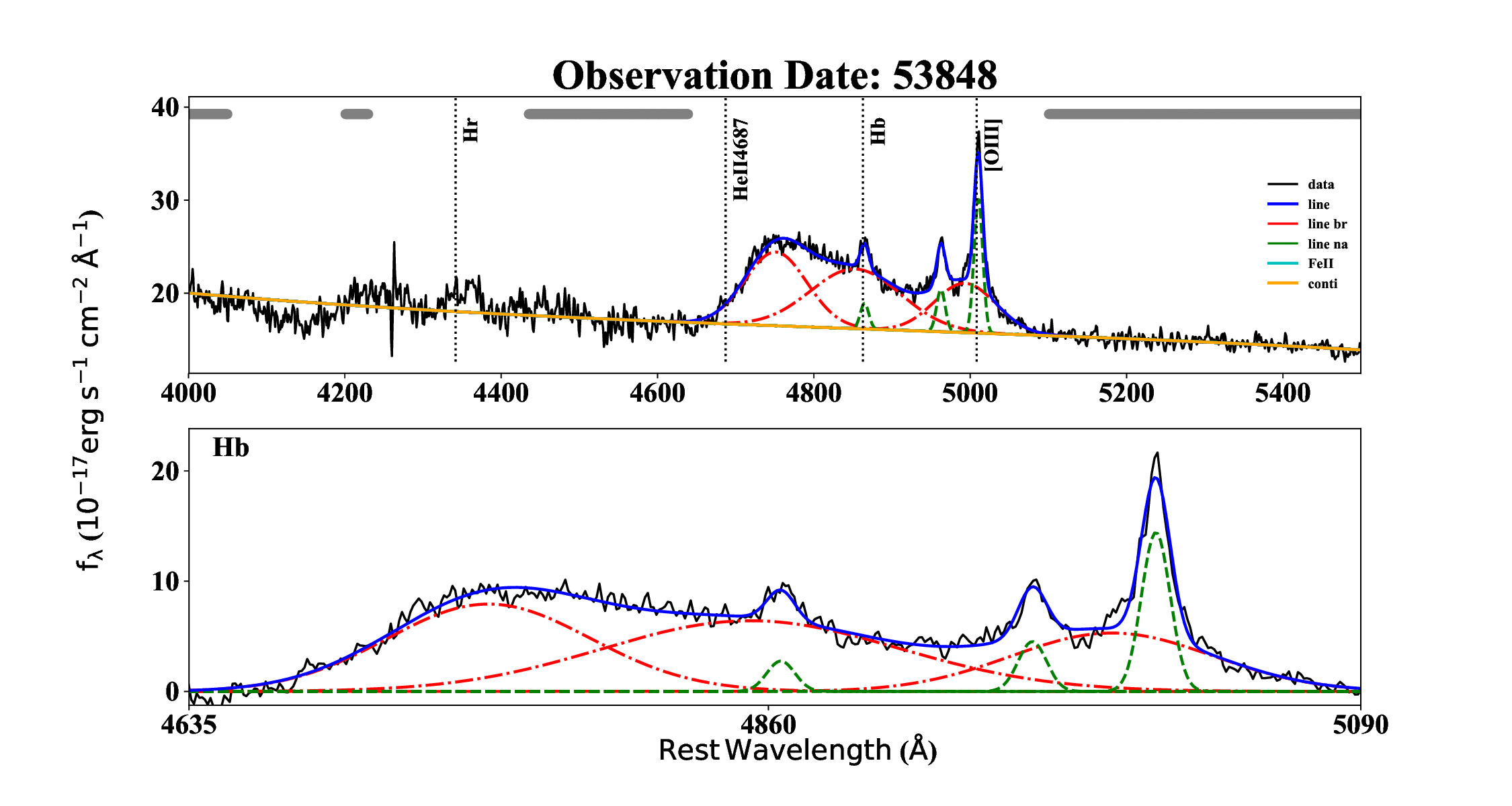}
\includegraphics[width=80mm, clip]{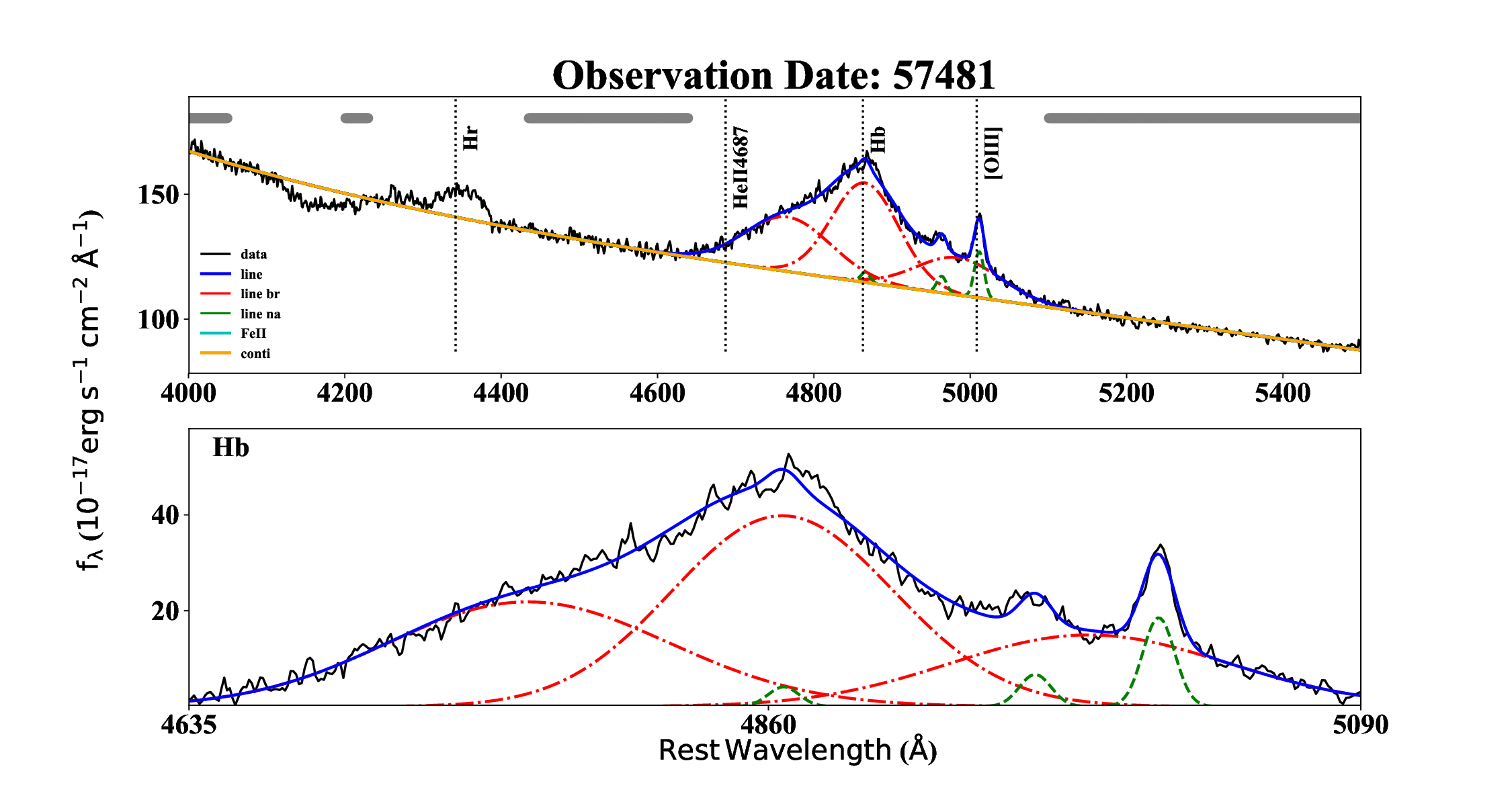}
\includegraphics[width=80mm, clip]{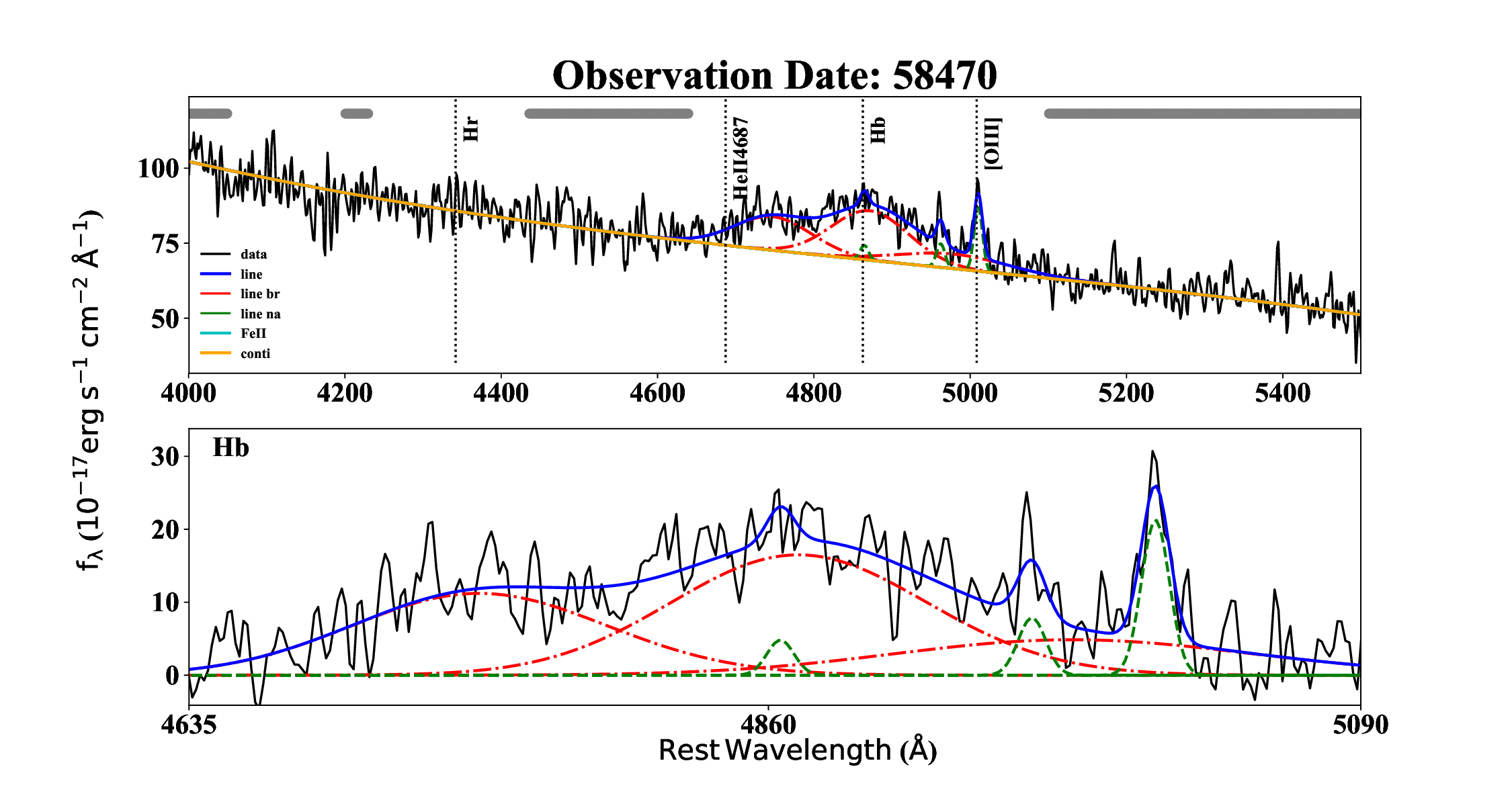}
\includegraphics[width=80mm, clip]{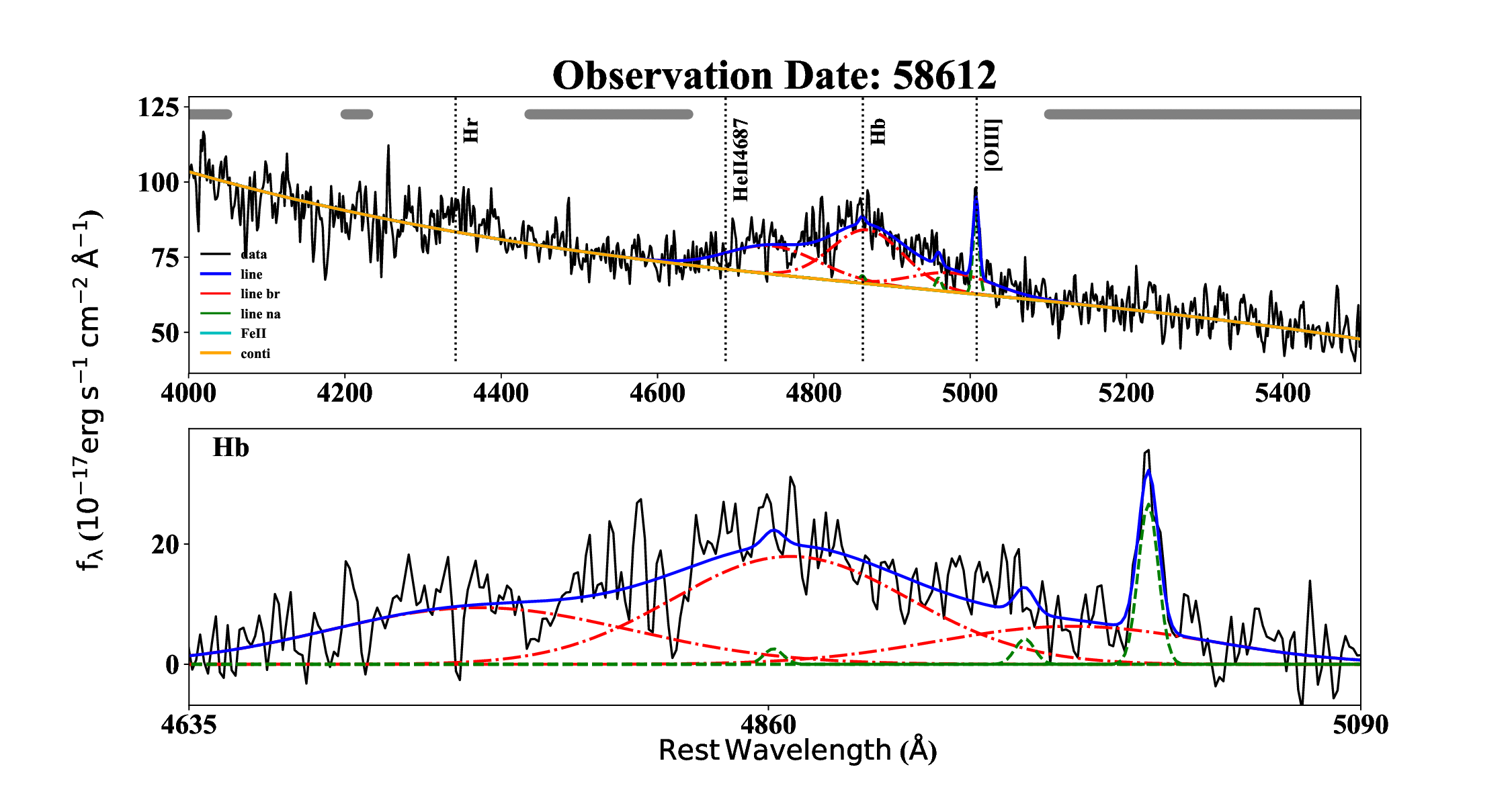}

\end{center}
\caption{
Result of the spectral fitting of the H$\rm{\beta}$ complex of SDSS J125809.31+351943.0. The upper two panels are for the SDSS spectra, and the lower two are for the MALLS spectra. The upper sub-panel of each panel shows the observed spectrum with the best-fit model, including the continuum whereas the lower sub-panel shows the H$\rm{\beta}$ and [OIII] lines after the best-fit continuum component is subtracted. Gaussians with dashed and dash-dotted lines represent narrow and broad line components, respectively.
}
\label{fig:fitting}
\end{figure}

The spectrum of this object is characterized by a wide, asymmetric H$\rm{\beta}$, which is blended with [OIII]. To estimate the H$\rm{\beta}$ flux we model-fitted the spectrum, where three Gaussians were used to approximate the broad asymmetric $\rm H\beta$ component, using the software ``PyQSOfit'' (\cite{2018ascl.soft09008G}). ``PyQSOfit'' fits the continuum component with a power-law functions and the emission lines with multiple Gaussian functions and estimates the errors of the model parameters with Markov Chain Monte Carlo simulations. Here, the galaxy components and iron lines are not taken into account because their contributions are small. In summary, the model components for the emission lines are three broad emission lines for H$\rm{\beta}$, one narrow line for H$\rm{\beta}$, two narrow lines for [OIII]4959/5007 (figure \ref{fig:fitting}). The parameters of the broad line component (center wavelength offset, line width, and scale factor) are allowed to vary with some fixed upper limits, whereas the center wavelength offset and line width of the narrow line component are fixed. The flux ratio between [OIII]5007 and narrow H$\rm{\beta}$ is determined from the fitting result of the SDSS spectrum with the best S/N ratio (MJD = 53848), the value of which is then assumed in the fitting of the other spectra ([OIII]5007 : H$\rm{\beta}$narrow = 3 : 0.61). For each spectrum, the flux is normalized by the average flux of [OIII]5007. 

\begin{figure}
 \begin{center}
 \includegraphics[width=80mm, clip]{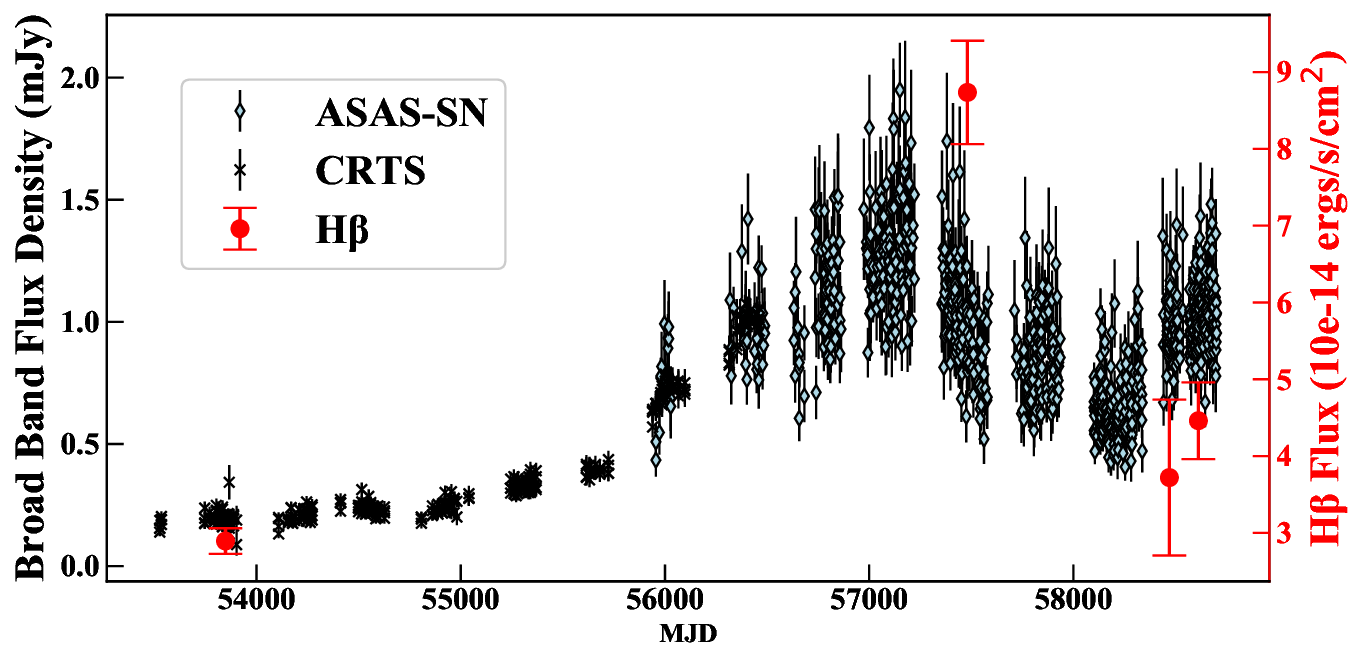}
 \end{center}
 \caption{
 Time-series of broad-band flux density and the H$\rm{\beta}$ flux of J1258. Crosses and diamonds denote the CRTS and ASAS-SN, respectively, in the scale indicated on the left vertical axis. Red circles denote the H$\rm{\beta}$ flux in the scale indicated on the right vertical axis.
 }
 \label{fig:curve_HB}
\end{figure}

Figure \ref{fig:curve_HB} shows the time-series of the total H$\rm{\beta}$ and continuum fluxes of this source. The ratio of H$\rm{\beta}$/[OIII] was found to have changed by 100\% during the observation period. Consequently, we identify J1258 as a new CSQ. Furthermore, the time-series shows that the source experienced two changing-states for the flux of H$\rm{\beta}$ and continuum flux density. We will discuss the variance of H$\rm{\beta}$ shape in a forthcoming paper on the basis of the on-going reverberation mapping of this object.

\section{Discussion} \label{sec:disussion}
J1258 shows a monotonic increase of luminosity for as much as 4 mag over about 30 years. This event is one of the largest amplitude of monotonic variations with the longest timescale of any reported quasar's variability so far. Since the mid-infrared luminosity varied corresponding to the optical luminosity, the variation of J1258 should be due to an intrinsic change of the mass accretion rate rather than absorption variation. It is also not likely to be a discrete event such as tidal disruption event or microlensing, as it is currently undergoing significant variations. We discuss here the 4 mag brightening event, which is the most significant feature of J1258.

In this chapter, we estimate the physical quantities of J1258 and discuss the mechanism of this event from three aspects: the emission line shape, the timescale, and the change in the Eddington ratio.

\subsection{The BEL shape}
One of the characteristic features of J1258 is its unique shape of BELs. In some reported CSQs, such unusual shapes of BELs exist (\cite{2019ApJ...874....8M}). Such a broad line profile is an interesting feature, suggesting a disk-shaped rotating BLR gas or a binary black hole. It is essential to discuss this shape's origin because we use the line width to later derive the black hole mass.

According to the disk-wind model of \cite{2014MNRAS.438.3340E}, the BLR is formed by the outflow from the accretion disk, and the shape of BEL is proposed to change in a sequence of steps depending on the parameter $L_{\rm BOL}/M^{2/3}$, where $L_{\rm BOL}$ is the bolometric luminosity, and $M$ is the black hole mass. When the luminosity is lower than a certain value, no gas is supplied to form the broad-line region; AGNs in this phase become ``true'' type-2 AGNs. When the luminosity rises, the weak outflow from the accretion disk forms a BLR near the surface of the accretion disk, creating double-peaked BELs.  When the luminosity rises more, the BEL distributes a wide region, including far from the accretion disk and changes to a single-peaked BELs (type-1 AGN). Comparing the shape of the H$\beta$ from the two SDSS spectra (Fig. \ref{fig:spectra}), the transition of this object's H$\beta$ shape is consistent with this model qualitatively, showing a single-peaked shape in its brighter phase and double-peaked shape in its fainter phase.

In the case of binary black holes, the emission lines can form various shapes depending on their masses and distances. One of the predicted features is that each component's velocity offset varies with the black holes' orbital motion (\cite{2014ApJ...789..140L}).

Volocity-resolved reverberation Mapping is an effective way to verify these scenarios. By measuring the time lag for each velocity component of the emission line, it is possible to detect outflow features (the time lag of the fast component should be smaller than the lag of the slow component). It is also possible to verify whether the velocity offset varies periodically based on the emission line shape's variation. 

The emission line of J1258 has three peaks, and their velocity offsets are almost symmetric. In order to reproduce this kind of emission line shape in a binary system, we have to take into account a complex system. Here, we assume the disk-wind model of \cite{2014MNRAS.438.3340E}, in which the spectra at the brightest period are closest to the type 1 AGN. In the next subsection, we estimate the physical quantities from the spectrum obtained when J1258 is closest to its brightest peak.

\subsection{Physical values of J1258}
Here, we estimate the bolometric luminosity, black hole mass, and Eddington ratio. For our calculations, we use the second SDSS spectrum (MJD=57481), which is the brightest one we have. 
First, we calculate the bolometric luminosity. To estimate from the optical spectrum, we adopt the formula for the bolometric correction in \cite{2019MNRAS.488.5185N}.
\begin{equation}
    L_{\rm BOL} = 40 \times \left( \frac{L(5100)}{10^{42}} \right) ^{-0.2} \times L(5100) 
\end{equation}
Here, $L_{\rm BOL}$ is the bolometric luminosity (erg/s), $L(5100)$ is the observed luminosity (erg/s) at 5100 {\AA} in the rest frame determined by power-law fitting of the spectrum. Substituting the $L(5100)$ obtained from the spectrum ($L(5100) = 10^{45.2} \ {\rm erg/s}$) into this equation, we derived $L_{BOL} = 10^{46.2} \ {\rm erg/s}$.

Next, we estimate the black hole masses. A black hole's virial mass can be estimated from from a single spectroscopic observation by assuming the formula empirically determined by the Reverberation mapping of nearby AGN. Here, we use the results of \cite{2006ApJ...641..689V} to calculate it. The formula is as follows.
\begin{eqnarray}
\label{eq:bhmass}
    {\rm log_{10}} \left( \frac{M_{\rm BH}}{M_{\odot}}\right) &=& 0.91 + 0.5 \times {\rm log_{10}}\left( \frac{L(5100)}{10^{44}}\right) \nonumber
    \\ &+& 2 \times {\rm log_{10}}({\rm FWHM} ({\rm H\beta}))
\end{eqnarray}
Here, $M_{\rm BH}$ is the black hole mass, and the ${\rm FWHM ( H\beta})$ is the full width at half maximum (km/s) of the broad H$\beta$ line. ${\rm FWHM (H\beta})$ is calculated from the spectra after subtracting the power-law continuum and the [OIII] component; ${\rm FWHM (H\beta}) = 1.03 \times 10^4 \ {\rm km/s}$.  As a result, the black hole mass was estimated to be $10^{9.55} \ M_{\odot}$. According to \cite{2006ApJ...641..689V}, the typical error of black hole masses derived from Eq. \ref{eq:bhmass} is about 0.5 dex. The Eddington limiting luminosity of this black hole mass is $L_{\rm EDD} = 10^{47.6} \ {\rm erg/s}$, giving the Eddington ratio of 0.04 in this epoch. Although there is no spectrum at the faintest time, we can infer that the Eddington ratio changed from about 0.001 to 0.04, given that the amplitude of the brightening event in optical is about factor 40. 

\subsection{Timescale}

We compare the duration of J1258's brightening event with the timescales of accretion disk variations. We consider four timescales; orbital timescale, thermal timescale, front propagation timescale, and viscous timescale (e.g., \cite{2018ApJ...864...27S}, \cite{2018MNRAS.480.3898N}). The orbital timescale ($t_{\rm orb}$) is the timescale on which the disk orbits. The thermal timescale ($t_{\rm th}$) is expressed as $t_{\rm th} = t_{\rm orb} / \alpha$ with the viscosity of $\alpha$. The cooling/heating front timescale ($t_{\rm front}$) is the radial propagation of the heating or cooling boundary across the disk. $t_{\rm front}$ is expressed as $t_{\rm front} = (H/R) ^{-1} t_{\rm th}$, where $H$ is the thickness of the disk and $R$ is the radius of the disk. The viscous timescale ($t_{\rm v}$) is the timescale that characterizes the mass flow, denoted by $t_{\rm v} = (H/R) ^{-2} t_{\rm th}$. \cite{2018ApJ...864...27S} estimated each of these timescales as follows.

\begin{eqnarray}
t_{\rm orb} &\sim& 10 \ {\rm days} \left( \frac{M_{\rm{BH}}}{10^8 M_{\odot}}\right) \left( \frac{R}{150R_{\rm g}}\right)^{3/2} \nonumber\\
t_{\rm th} 
&\sim& 1 \ {\rm yr} 
\left( \frac{\alpha}{0.03} \right) ^ {-1}
\left( \frac{M_{\rm BH}}{10^8 M_\odot} \right) 
\left( \frac{R}{150 R_{\rm g}} \right)^{3/2} \nonumber\\
t_{\rm front} 
&\sim& 20 \ {\rm yr}
\left( \frac{H/R}{0.05} \right) ^ {-1}
\left( \frac{\alpha}{0.03} \right) ^ {-1}
\left( \frac{M_{\rm BH}}{10^8 M_\odot} \right)
\left( \frac{R}{150 R_{\rm g}} \right)^{3/2} \nonumber\\
t_{\rm v}
&\sim& 400 \ {\rm yr}
\left( \frac{H/R}{0.05} \right) ^ {-2}
\left( \frac{\alpha}{0.03} \right) ^ {-1}
\left( \frac{M_{\rm BH}}{10^8 M_\odot} \right)
\left( \frac{R}{150 R_{\rm g}} \right)^{3/2} \nonumber
\end{eqnarray}
$R_{\rm g}$ is the gravity radius defined as $GM_{\rm BH}/c^2$. 
We calculated the above four timescales for J1258, assuming that radiation in optical (typical temperature is $10^{3-4}$ K) is dominant at about $150R_{\rm g}$ in the standard disk model. Then, each timescale is calculated as follows:
$t_{\rm orb} \sim 300 \ {\rm days},\ 
t_{\rm th} \sim 30 \ {\rm years},\ 
t_{\rm front} \sim 600 \ {\rm years},\ 
t_{\rm v} \sim 12,000 \ {\rm years}$.
The results indicate that 30 years is the closest to the thermal timescale. 

On the other hand, \cite{2018ApJ...864...27S} points out that substituting other plausible values for $\alpha$ and $H/R$ can significantly change these values. The disk is likely to heat up and inflate to $H/R \sim 0.1$ without the assumption in the standard thin disk model (\cite{1973A&A....24..337S}) of no torque at the inner edge of the accretion disk (e.g., \cite{2000ApJ...528..161A}). Also, \cite{2007MNRAS.376.1740K} argues that the viscosity $\alpha$ is likely to be $\sim$ 0.3 from observations. In that case, the timescale is rewritten as follows:
$t_{\rm orb} \sim 300 \ {\rm days},\ 
t_{\rm th} \sim 3 \ {\rm years},\ 
t_{\rm front} \sim 30 \ {\rm years},\ 
t_{\rm v} \sim 300 \ {\rm years}$.

Thus, this phenomenon can be explained by thermal process or heating front propagation in terms of timescales (\cite{2018MNRAS.480.4468R}). A situation in which a thermal process induces a brightening phenomenon is that the entire disk is heated by high energy radiation from a small central region (\cite{2014ApJ...788...48S}, \cite{2019ApJ...885...44D}). However, we consider this situation is unlikely to occur because the H$\beta$ variation lags behind the continuum variation (see Fig. \ref{fig:curve_HB}), suggesting that the BLR is ionized by the accretion disk rather than the high energy radiation from the central region.

\subsection{Eddington ratio}
\cite{2018MNRAS.480.3898N} explained extremely variable AGNs with a disk state transition (switch between low/hard state to high/soft state) and thermal front propagation. They divided extremely variable AGNs into three categories:
\begin{enumerate}
\item One showing a state change due to disc evaporation/condensation associated with a factor 2–4 decrease/increase in luminosity
\item One with mass accretion rate change due to the thermal front propagation
\item One showing both
\end{enumerate}
(1) and (3) accompany a changing-look phenomenon, a significant change of BEL, but (2) does not. In other words, the brightening phenomenon of J1258, which shows a 40-fold brightening and a significant change in H$\beta$ intensity, is classified as (3). Here, \cite{2018MNRAS.480.3898N} predicted that the Eddington ratio crosses the border about $\sim 0.01$ when a state transition like (1) or (3) happens. This prediction is confirmed in J1258 (the Eddington ratio changed from 0.001 to 0.04).

\subsection{Comparison to other variable quasars}
Most of the CSQs reported so far have had timescales of less than ten years and amplitudes of optical variability with less than factor 10 (e.g., \cite{2019ApJ...870...65I}). The amount of variability is roughly proportional to the accretion disk area with the cooling/heating front propagates (\cite{2019ApJ...874....8M}). The size of the accretion disk, $t_{\rm th}$, and $t_{\rm front}$ are values that correlate the black hole mass. The black hole mass of J1258 is more than ten times larger than that of the previously reported CSQ black hole masses, which are typically about $10^{7 - 8}$ solar masses (e.g., \cite{2016MNRAS.457..389M}). This suggests that the previously reported brightening phenomena of CSQ and J1258 can be interpreted as the same in terms of timescales and amplitudes. We expect that quasars with larger amplitudes of variability will be discovered in samples with larger black hole mass.

\subsection{Comparison with \cite{2020MNRAS.491.4925G}} \label{comparison}
The quasar catalog used in this study, SDSS DR7Q, is included in the Million Quasar Catalog \footnote{http://quasars.org/milliquas.htm} v5.2 used in \cite{2020MNRAS.491.4925G}. In both studies, CSQ samples were selected using the CRTS light curve. The main difference in the selection method is the analysis method of the light curve.

\begin{figure}
\begin{center}
\includegraphics[width=80mm, clip]{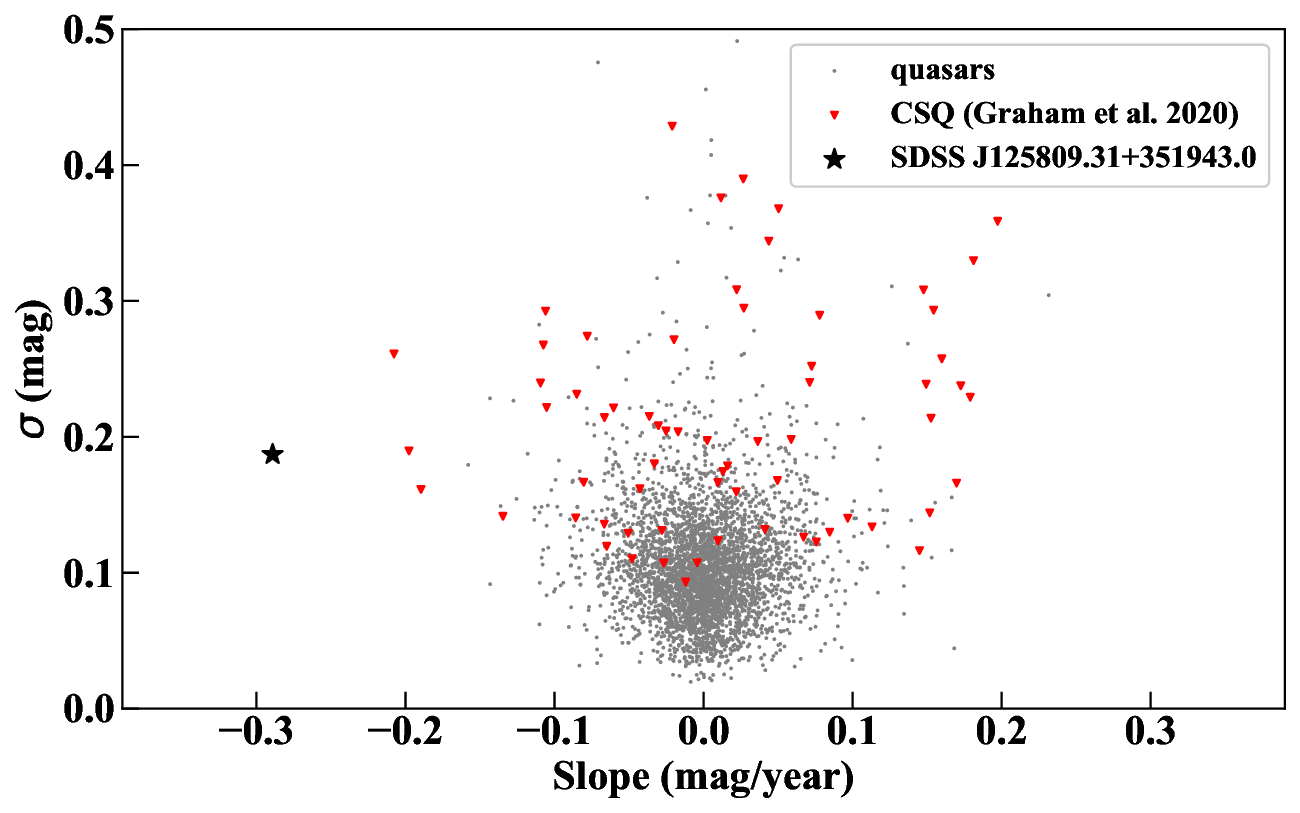}
\end{center}
\caption{Same as Figure \ref{fig:slope_sigma}, but the CSQ sample of \cite{2020MNRAS.491.4925G} included in SDSS DR7Q is shown as red triangle.}
\label{fig:slope_sigma_graham}
\end{figure}

In our method, the full ranges of the light curves are fitted with linear functions. In \cite{2020MNRAS.491.4925G}, they use Bayesian Blocks to estimate the amount of variability. The advantage of The advantage of the linear fitting is that it is sensitive to monotonic variations with timescales, which is longer than the light curve's full range. However, the disadvantage is that it misses the variations in various timescales between the light curves. The Bayesian Block, on the other hand, is capable of detecting any timescale variations that occur within the light curves. In figure \ref{fig:slope_sigma_graham}, we plot how the CSQs listed in \cite{2020MNRAS.491.4925G} are distributed in the ``slope-sigma plan'' of our linear fitting. In fact, the results show that CSQs are distributed unevenly where the sigma is larger than 0.1 because they are recognized as scatters when they change on a timescale shorter than the interval of the light curve.

In the case of J1258, it was detectable by fitting with a linear function because of its monotonic brightening during all range of the light curve. However, such an object should also be detected by using the Bayesian Block. A possible reason is the indeterminacy of the emission line flux measurement on the unusual shape of the BELs; in \cite{2020MNRAS.491.4925G}, H$\beta$ is fitted with two Gaussians to obtain the flux. Therefore, in the case of H$\beta$ with an unusual shape, the flux measurement may be significantly affected.

\section{Conclusions} \label{sec:conclusion}
We discovered an extremely variable quasar, SDSS J125809.31+351943.0, brightened for 4 mag from 1983 to 2015. We identified this object as a new CSQ on the basis of the significant changes in the mid-infrared luminosity and in the intensity of the broad emission line. The weakness of the radio emission indicates that the variability is not caused by the jet. Furthermore, the flux in mid-infrared changed following the optical, which suggests that the accretion disk itself changed, rather than the variation of absorption. The object is characterized by an unusual shape of H$\rm{\beta}$, and variability of 4 mag. The shape of H$\rm{\beta}$ is consistent with the disk-wind model by \cite{2014MNRAS.438.3340E}. The variability of 4 mag is consistent with other variable quasars in terms of timescales and amplitude, except for the black hole mass of $\sim$ 10 times larger than other reported CSQs. This object is one of the most drastically variable objects so far, which will give us a new insight into accretion physics. To investigate further, a project of reverberation mapping of this object using 3.8-m Seimei Telescope in Japan is currently underway.

\begin{ack}
We thank Tomohito Ohshima, Satoshi Honda, Yohichi Itoh and Yuhei Takagi for their help in writing the proposal to obtain the observation time used in this study. 

The CSS survey is funded by the National Aeronautics and Space
Administration under Grant No. NNG05GF22G issued through the Science
Mission Directorate Near-Earth Objects Observations Program.  The CRTS
survey is supported by the U.S.~National Science Foundation under
grants AST-0909182 and AST-1313422.

Funding for the Sloan Digital Sky Survey IV has been provided by the Alfred P. Sloan Foundation, the U.S. Department of Energy Office of Science, and the Participating Institutions. SDSS-IV acknowledges
support and resources from the Center for High-Performance Computing at
the University of Utah. The SDSS web site is www.sdss.org.

SDSS-IV is managed by the Astrophysical Research Consortium for the 
Participating Institutions of the SDSS Collaboration including the 
Brazilian Participation Group, the Carnegie Institution for Science, 
Carnegie Mellon University, the Chilean Participation Group, the French Participation Group, Harvard-Smithsonian Center for Astrophysics, 
Instituto de Astrof\'isica de Canarias, The Johns Hopkins University, Kavli Institute for the Physics and Mathematics of the Universe (IPMU) / 
University of Tokyo, the Korean Participation Group, Lawrence Berkeley National Laboratory, 
Leibniz Institut f\"ur Astrophysik Potsdam (AIP),  
Max-Planck-Institut f\"ur Astronomie (MPIA Heidelberg), 
Max-Planck-Institut f\"ur Astrophysik (MPA Garching), 
Max-Planck-Institut f\"ur Extraterrestrische Physik (MPE), 
National Astronomical Observatories of China, New Mexico State University, 
New York University, University of Notre Dame, 
Observat\'ario Nacional / MCTI, The Ohio State University, 
Pennsylvania State University, Shanghai Astronomical Observatory, 
United Kingdom Participation Group,
Universidad Nacional Aut\'onoma de M\'exico, University of Arizona, 
University of Colorado Boulder, University of Oxford, University of Portsmouth, 
University of Utah, University of Virginia, University of Washington, University of Wisconsin, 
Vanderbilt University, and Yale University.

This publication makes use of data products from the Wide-field Infrared Survey Explorer, which is a joint project of the University of California, Los Angeles, and the Jet Propulsion Laboratory/California Institute of Technology, funded by the National Aeronautics and Space Administration.
\end{ack}
\bibliography{main.bib}{}

\begin{thebibliography}{}

\bibitem[{Agol} and {Krolik}(2000)]{2000ApJ...528..161A}
  {Agol}, Eric, \& {Krolik}, Julian~H.\ 2000, \apj, 528, 161

\bibitem[{Alam} et~al.(2015)]{2015ApJS..219...12A}
  {Alam}, Shadab, {et~al.}\ 2015, \apjs, 219, 12

\bibitem[{Becker} et~al.(1995)]{1995ApJ...450..559B}
  {Becker}, Robert~H., {White}, Richard~L., \& {Helfand}, David~J.\ 1995, \apj,
  450, 559

\bibitem[{Bruce} et~al.(2017)]{2017MNRAS.467.1259B}
  {Bruce}, A., {et~al.}\ 2017, \mnras, 467, 1259

\bibitem[{Clavel} et~al.(1989)]{1989ApJ...337..236C}
  {Clavel}, J., {Wamsteker}, W., \& {Glass}, I.~S.\ 1989, \apj, 337, 236

\bibitem[{Cohen} et~al.(1986)]{1986ApJ...311..135C}
  {Cohen}, Ross~D., {Rudy}, Richard~J., {Puetter}, R.~C., {Ake}, T.~B., \&
  {Foltz}, Craig~B.\ 1986, \apj, 311, 135

\bibitem[{Dexter} et~al.(2019)]{2019ApJ...885...44D}
  {Dexter}, Jason, {et~al.}\ 2019, \apj, 885, 44

\bibitem[{Drake} et~al.(2019)]{2019MNRAS.482...98D}
  {Drake}, A.~J., {Djorgovski}, S.~G., {Graham}, M.~J., {Stern}, D., {Mahabal},
  A.~A., {Catelan}, M., {Christensen}, E., \& {Larson}, S.\ 2019, \mnras, 482,
  98

\bibitem[{Drake} et~al.(2009)]{2009ApJ...696..870D}
  {Drake}, A.~J., {et~al.}\ 2009, \apj, 696, 870

\bibitem[{Elitzur}(2012)]{2012ApJ...747L..33E}
  {Elitzur}, Moshe\ 2012, \apjl, 747, L33

\bibitem[{Elitzur} et~al.(2014)]{2014MNRAS.438.3340E}
  {Elitzur}, Moshe, {Ho}, Luis~C., \& {Trump}, Jonathan~R.\ 2014, \mnras, 438,
  3340

\bibitem[{Gezari} et~al.(2017)]{2017ApJ...835..144G}
  {Gezari}, S., {et~al.}\ 2017, \apj, 835, 144

\bibitem[{Graham} et~al.(2020)]{2020MNRAS.491.4925G}
  {Graham}, Matthew~J., {et~al.}\ 2020, \mnras, 491, 4925

\bibitem[{Gunn} et~al.(2006)]{2006AJ....131.2332G}
  {Gunn}, James~E., {et~al.}\ 2006, \aj, 131, 2332

\bibitem[{Guo} et~al.(2020)]{2020ApJ...888...58G}
  {Guo}, Hengxiao, {et~al.}\ 2020, \apj, 888, 58

\bibitem[{Guo} et~al.(2018)]{2018ascl.soft09008G}
  {Guo}, Hengxiao, {Shen}, Yue, \& {Wang}, Shu, 2018, {PyQSOFit: Python code to
  fit the spectrum of quasars}

\bibitem[{Husemann} et~al.(2016)]{2016A&A...593L...9H}
  {Husemann}, B., {et~al.}\ 2016, \aap, 593, L9

\bibitem[{Hutsem{\'e}kers} et~al.(2017)]{2017A&A...604L...3H}
  {Hutsem{\'e}kers}, D., {Ag{\'\i}s Gonz{\'a}lez}, B., {Sluse}, D., {Ramos
  Almeida}, C., \& {Acosta Pulido}, J.~A.\ 2017, \aap, 604, L3

\bibitem[{Ichikawa} et~al.(2019)]{2019ApJ...870...65I}
  {Ichikawa}, Kohei, {Ueda}, Junko, {Bae}, Hyun-Jin, {Kawamuro}, Taiki,
  {Matsuoka}, Kenta, {Toba}, Yoshiki, \& {Shidatsu}, Megumi\ 2019, \apj, 870,
  65

\bibitem[{Kellermann} et~al.(1989)]{1989AJ.....98.1195K}
  {Kellermann}, K.~I., {Sramek}, R., {Schmidt}, M., {Shaffer}, D.~B., \&
  {Green}, R.\ 1989, \aj, 98, 1195

\bibitem[{King} et~al.(2007)]{2007MNRAS.376.1740K}
  {King}, A.~R., {Pringle}, J.~E., \& {Livio}, M.\ 2007, \mnras, 376, 1740

\bibitem[{Kochanek} et~al.(2017)]{2017PASP..129j4502K}
  {Kochanek}, C.~S., {et~al.}\ 2017, \pasp, 129, 104502

\bibitem[{LaMassa} et~al.(2015)]{2015ApJ...800..144L}
  {LaMassa}, Stephanie~M., {et~al.}\ 2015, \apj, 800, 144

\bibitem[{Liu} et~al.(2014)]{2014ApJ...789..140L}
  {Liu}, Xin, {Shen}, Yue, {Bian}, Fuyan, {Loeb}, Abraham, \& {Tremaine},
  Scott\ 2014, \apj, 789, 140

\bibitem[{MacLeod} et~al.(2010)]{2010ApJ...721.1014M}
  {MacLeod}, C.~L., {et~al.}\ 2010, \apj, 721, 1014

\bibitem[{MacLeod} et~al.(2019)]{2019ApJ...874....8M}
  {MacLeod}, Chelsea~L., {et~al.}\ 2019, \apj, 874, 8

\bibitem[{MacLeod} et~al.(2016)]{2016MNRAS.457..389M}
  {MacLeod}, Chelsea~L., {et~al.}\ 2016, \mnras, 457, 389

\bibitem[{Mainzer} et~al.(2014)]{2014ApJ...792...30M}
  {Mainzer}, A., {et~al.}\ 2014, \apj, 792, 30

\bibitem[{Merloni} et~al.(2015)]{2015MNRAS.452...69M}
  {Merloni}, A., {et~al.}\ 2015, \mnras, 452, 69

\bibitem[{Netzer}(2019)]{2019MNRAS.488.5185N}
  {Netzer}, Hagai\ 2019, \mnras, 488, 5185

\bibitem[{Noda} and {Done}(2018)]{2018MNRAS.480.3898N}
  {Noda}, Hirofumi, \& {Done}, Chris\ 2018, \mnras, 480, 3898

\bibitem[{Pancoast} et~al.(2014)]{2014MNRAS.445.3073P}
  {Pancoast}, Anna, {Brewer}, Brendon~J., {Treu}, Tommaso, {Park}, Daeseong,
  {Barth}, Aaron~J., {Bentz}, Misty~C., \& {Woo}, Jong-Hak\ 2014, \mnras, 445,
  3073

\bibitem[{Peterson} et~al.(2004)]{2004ApJ...613..682P}
  {Peterson}, B.~M., {et~al.}\ 2004, \apj, 613, 682

\bibitem[{Ross} et~al.(2018)]{2018MNRAS.480.4468R}
  {Ross}, Nicholas~P., {et~al.}\ 2018, \mnras, 480, 4468

\bibitem[{Ruan} et~al.(2016)]{2016ApJ...826..188R}
  {Ruan}, John~J., {et~al.}\ 2016, \apj, 826, 188

\bibitem[{Rumbaugh} et~al.(2018)]{2018yCat..18540160R}
  {Rumbaugh}, N., {et~al.}\ 2018, VizieR Online Data Catalog,  J/ApJ/854/160

\bibitem[{Schneider} et~al.(2010)]{2010AJ....139.2360S}
  {Schneider}, Donald~P., {et~al.}\ 2010, \aj, 139, 2360

\bibitem[{Shakura} and {Sunyaev}(1973)]{1973A&A....24..337S}
  {Shakura}, N.~I., \& {Sunyaev}, R.~A.\ 1973, \aap, 500, 33

\bibitem[{Shappee} et~al.(2014)]{2014ApJ...788...48S}
  {Shappee}, B.~J., {et~al.}\ 2014, \apj, 788, 48

\bibitem[{Sheng} et~al.(2019)]{2019arXiv190502904S}
  {Sheng}, Zhenfeng, {et~al.}\ 2019, arXiv e-prints,  arXiv:1905.02904

\bibitem[{Stern} et~al.(2018)]{2018ApJ...864...27S}
  {Stern}, Daniel, {et~al.}\ 2018, \apj, 864, 27

\bibitem[{Storchi-Bergmann} et~al.(1995)]{1995ApJ...443..617S}
  {Storchi-Bergmann}, Thaisa, {Eracleous}, Michael, {Livio}, Mario, {Wilson},
  Andrew~S., {Filippenko}, Alexei~V., \& {Halpern}, Jules~P.\ 1995, \apj, 443,
  617

\bibitem[{Tohline} and {Osterbrock}(1976)]{1976ApJ...210L.117T}
  {Tohline}, J.~E., \& {Osterbrock}, D.~E.\ 1976, \apjl, 210, L117

\bibitem[{Trakhtenbrot} et~al.(2019)]{2019ApJ...883...94T}
  {Trakhtenbrot}, Benny, {et~al.}\ 2019, \apj, 883, 94

\bibitem[{Vestergaard} and {Peterson}(2006)]{2006ApJ...641..689V}
  {Vestergaard}, Marianne, \& {Peterson}, Bradley~M.\ 2006, \apj, 641, 689

\bibitem[{Wang} et~al.(2018)]{2018ApJ...858...49W}
  {Wang}, J., {Xu}, D.~W., \& {Wei}, J.~Y.\ 2018, \apj, 858, 49

\bibitem[{Weedman}(1985)]{1985ApJS...57..523W}
  {Weedman}, D.~W.\ 1985, \apjs, 57, 523

\bibitem[{Wright} et~al.(2010)]{2010AJ....140.1868W}
  {Wright}, Edward~L., {et~al.}\ 2010, \aj, 140, 1868

\bibitem[{York} et~al.(2000)]{2000AJ....120.1579Y}
  {York}, Donald~G., {et~al.}\ 2000, \aj, 120, 1579

\end{thebibliography}
\bibliographystyle{main.bst}

\end{document}